\documentclass[aps,pre,preprint]{revtex4-2}
\usepackage{amsmath}
\usepackage{graphicx}
\usepackage{bbm}
\usepackage{subfigure}
\usepackage[caption=false]{subfig}
\begin{document}
\title{Energy Band Structure of Relativistic Quantum Plasmon Excitation}
\author{M. Akbari-Moghanjoughi}
\affiliation{Faculty of Sciences, Department of Physics, Azarbaijan Shahid Madani University, 51745-406 Tabriz, Iran\\ massoud2002@yahoo.com}

\begin{abstract}
In this paper we use the effective Schr\"{o}dinger-Poisson and square-root Klein-Gordon-Poisson models to study the quantum and relativistic quantum energy band structure of finite temperature electron gas in a neutralizing charge background. Based the plasmon band gap appearing above the Fermi level, new definitions on plasmonic excitations and plasma parameters in a wide electron temperature-density regime is suggested. The new equation of state (EoS) for excited electrons to the plasmon band leads to novel aspects of relativistic collective quantum excitations such as the plasmon black-out and quantum pressure collapse which are studied using both non-relativistic and relativistic quantum models. The plasmon black-out effect may be used to explain why metallic elements do not show collective behavior at low temperatures. The model can be used to predict phases of matter in which the plasmonic activities is shut down, hence, it may behave like a mysterious dark matter. On the other hand, the energy band structure model predicts the plasmon pressure collapse in temperature-density coordinates matching that of a white dwarf star. The prediction of energy band structure of collective quantum excitations may have direct implications for the inertial confinement fusion (ICF), the EoS of warm dense matter (WDM) and evolution of stellar and other unknown cosmological structures. It is found that predictions of non-relativistic and relativistic quantum excitation models closely match up to temperature-density of degenerate stars which confirms the relevance of non-relativistic plasmon models used in the warm and dense matter regime. The effect of positron on band structure of collective quantum excitations is also studied.
\end{abstract}

\date{\today}

\maketitle
\newpage

\section{Introduction}

The technology is rapidly moving towards the plasmonics and nanoplasmonics designs \cite{atw,maier,man1} which provide efficient and more reliable methods of communication and energy transport. The terahertz scale response of electron oscillations, makes the plasmonic an ideal candidate for fast electronic switches in integrated circuits \cite{hu1,seeg}. Also, due to collective effects in plasmon excitations a large number of electrons contribute to quantum phenomena instead of single electron-hole mechanisms in ordinary semiconductor devices \cite{kit,ash}. Plasmonics has become an active field of interdisciplinary research, since mid 19's after realization of the surface plasmon-polariton resonance effect and the surface-enhanced Raman scattering (SERS) effect in 1970's \cite{xu}. In plasmonic devices collective oscillations of free electrons is driven by either electromagnetic radiation on appropriate plasmonic material causing the localised surface plasmon resonance (LSPR) \cite{may} or by external field stimulation coupled to the device causing the hot electron ejection which then are collected in a Schottky junction with an appropriate semiconductor coating. Due to extreme sensitivity of plasmonic devices to the size and geometry a wide range of applications are foreseeable in this technology \cite{yofee,zhu}. However, the efficient plasmonic designs is twisted with many technological obstacles which slows down the rapid developments, such as that occurred for semiconductor industry \cite{lucio,mark,haug,gardner}. Plasmon effects dominate mostly at nanoscale which is relatively costly to fabricate. By overcoming these limitations, plasmonic devices can find their full potential in fields such as plasmon focusing \cite{stock}, optical emitters \cite{and}, solar cells \cite{zhuo}, nanoscale waveguiding \cite{qui}, optical antennas \cite{muh}, communication devices \cite{umm}, plasmonic sensors \cite{goy} and modulator \cite{haf} nanoscale swiches \cite{kar} and even spasers \cite{oul}. The plasmonic energy conversion device may require few important technological considerations \cite{jian,fey1,hugen,sun1,sun2,sun3,cesar} with respect to the materials, integration and low dimensional fabrication and final assembly \cite{jac}. Investigations have shown that harvesting of the electromagnetic energy depends on the right choice of plasmonic geometries and fine-tuned design of nanoplasmonic device in order to make more effective solar-cell technology. In this way the solar energy is efficiently transferred to the large collection of electrons and can be absorbed in a single quantum well, in array of quantum dots or molecular chromophores \cite{at2}.

Collective electron excitations play inevitable role in astrophysical plasmas \cite{salpeter,hoyos,chandra2,kothary}, on the other hand. They contribute to many important linear and nonlinear effects in laboratory and cosmological scales, such as the charge screening, communication black-out, collisionless Landau damping, large variety of instabilities, localized density formations as soliton and shock waves, harmonic generation and endless other phenomena \cite{chen}. In extreme temperature and density conditions, collective electron effects combine with quantum and relativity to produce some novel phenomena such as stellar pressure collapse beyond the Chandrasekhar mass limit \cite{chandra3,chandra}. In the laboratory, the inertial confinement fusion (ICF) \cite{krall} device uses a delicate design to compress and heat the thermonuclear fuel containing palette in order to ignite the efficient nuclear fusion towards an efficient and clean source of energy. Benefits of numerous plasma applications are indebted to the theoretical and experimental developments which take place over the past century with pioneering works of many people \cite{fermi,bohm,bohm1,bohm2,pines,levine,klim,5,6,7,8,9,50,51,52,53,lind}. There has been a large attempt toward developing efficient tools to cope with many-body collective effects in quantum plasmas in recent decade. From Wigner-Poisson-Maxwell kinetic theory and corresponding quantum hydrodynamic developments \cite{haas1,man2,man3,haasbook} to density functional theory \cite{axel,fischer} and quantum Monte Carlo technique are all effective tool with their pros and cones. Due to simplicity, quantum hydrodynamic theory has attracted increased attraction over the past few years \cite{se,sten,ses,brod1,mark1,man4,scripta,stenf1,stenf2,stenf3,stenf4,ionst}. However, due to dual-scale nature of electron oscillations in quantum plasmas which was not incorporated in the original quantum hydrodynamic theory some results has led to intense recent debate among researchers \cite{bonitz1,sea1,bonitz2,sea2,bonitz3,akbarihd,sm,michta,moldabekov}. On the other hand, the effective Schr\"{o}dinger-Poisson model \cite{fhaas} has shown some recent success in capturing essential features of collective quantum excitations \cite{akbquant,akbheat,akbdual,akbedge,akbint,akbnat1,akbnat2}. The latter model suggests that collective quasiparticles in a dense quantum plasma can behave as if they possess two distinct de Broglie's wavelengths corresponding to wave-like and particle-like plasmon oscillation. It has been shown that this fundamental aspect leads to new features of collective quantum excitations. In current work we use the square-root Klein-Gordon-Poisson model \cite{fhaas2} to deduce the energy band structure of relativistic quantum plasmon excitations which extends our previous work for applications in a wide range of density and temperature, including astrophysical plasmas. Within the relativistic quantum band structure model we study various thermodynamic quantities of plasmon excitations in extreme density-temperature regimes, where, the non-relativistic theories can not be used.

\section{Collective Quantum Electron Excitations}

The collective electron excitations in a non-relativistic electron gas of arbitrary degenerate electron gas with a positive neutralizing background is modeled via the following coupled effective Schr\"{o}dinger-Poisson system \cite{fhaas,akbedge}
\begin{subequations}\label{sp}
\begin{align}
&i\hbar \frac{{\partial {\cal N}({\bf{r}},t)}}{{\partial t}} =  - \frac{{{\hbar ^2}}}{{2m}}\Delta {\cal N}({\bf{r}},t) - e\phi ({\bf{r}}){\cal N}({\bf{r}},t) + \mu {\cal N}({\bf{r}},t),\\
&\Delta \phi ({\bf{r}}) = 4\pi e\left[ {|{\cal N}({\bf{r}}){|^2} - {n_0}} \right],
\end{align}
\end{subequations}
where ${\cal{N}}({\bf{r}},t)=\psi({\bf{r}},t)\exp[iS({\bf{r}},t)/\hbar]$ characterizes the statefunction with $n({\bf{r}})=\psi({\bf{r}})\psi^*({\bf{r}})$ being the local electron number density and ${{p(r,t) = }}\nabla {{S(r,t)}}$ being the electron momentum. The electrostatic interaction between electron is modeled via the scalar potential $\phi({\bf{r}})$ which couples the Schr\"{o}dinger equation to the Poisson relation and $\mu$ is the chemical potential which is defined through the following non-relativistic isothermal equation of state (EoS)
\begin{subequations}\label{np}
\begin{align}
&{n_e(\mu,T)} = \frac{{{2^{1/2}}m{^{3/2}}}}{{{\pi ^2}{\hbar ^3}}}  \int_{0}^{ + \infty } {\frac{{\sqrt{{\varepsilon}} d{\varepsilon}}}{{{e^{\beta ({\varepsilon}-\mu)}} + 1}}},\\
&{P_e(\mu,T)} = \frac{{{2^{3/2}} m{^{3/2}}}}{{3{\pi ^2}{\hbar ^3}}}\int_0^{ + \infty } {\frac{{{{\varepsilon}^{3/2}} d{\varepsilon}}}{{{e^{\beta ({\varepsilon} - {\mu})}} + 1}}.}
\end{align}
\end{subequations}
where $\beta=1/k_B T$ with $T$ being the equilibrium electron temperature and $P_e$ being the quantum statistical electron gas pressure. The simple thermodynamic relation, $n_e\nabla\mu=\nabla P_e(n_e)$, holds between the dependent thermodynamic quantities. In the quasi-stationary limit $p=0$, the statefunction modulus may be decomposed into the separate variable $\psi({\bf{r}},t)=\psi(t)\psi({\bf{r}})$ functionals and one arrives at the following linear coupled pseudoforce system
\begin{subequations}\label{pf}
\begin{align}
&i\hbar \frac{{d\psi(t)}}{{dt}} = \varepsilon\psi(t),\\
&\Delta \Psi ({\bf{r}}) + \Phi ({\bf{r}}) + E\Psi ({\bf{r}}) = 0,\\
&\Delta \Phi ({\bf{r}}) - \Gamma \Psi ({\bf{r}}) = 0,
\end{align}
\end{subequations}
where we have used the expansion scheme $\{\psi^0=1,\phi^0=0,\mu^0=\mu_0\}$ and used the normalized functionals $\Psi({\bf{r}})=\psi({\bf{r}})/n_0$ with $n_0$ being the equilibrium electron number density, $\Phi({\bf{r}})=e\phi({\bf{r}})/E_0$ with $E_0=m_0 c^2$ being the electron rest energy. The parameter $\Gamma=8\alpha R^3/3\pi$ characterized the collective electrostatic interaction strength with $R=(n_0/n_c)^{1/3}$ being the relativity parameter and $n_c=k_c^3/3\pi^2$ being the characteristic Compton number density defined through the Compton wavenumber $k_c=m_0 c/\hbar$. Note that the number density $n_c$ defines a single electron inside the Compton sphere of radius equal to the Compton wavelength $\lambda_c=h/m_0 c$. The reason for normalization in relativistic units is the later comparison between non-relativistic and the fully relativistic quantum electron gas models. The normalized energy $E=(\epsilon-\mu_0)/E_0$ is the kinetic of electrons as measured from top of the electron Fermi sea, which in this case, is $\mu_0$ and depends on both temperature and density of the arbitrary degenerate electron gas. In this normalization the space and time variables are then normalized to the Compton wavelength and characteristic Compton frequency, $\omega_c=\hbar k_c^2/2 m_0$, respectively.

\begin{figure}[ptb]\label{Figure1}
\includegraphics[scale=0.57]{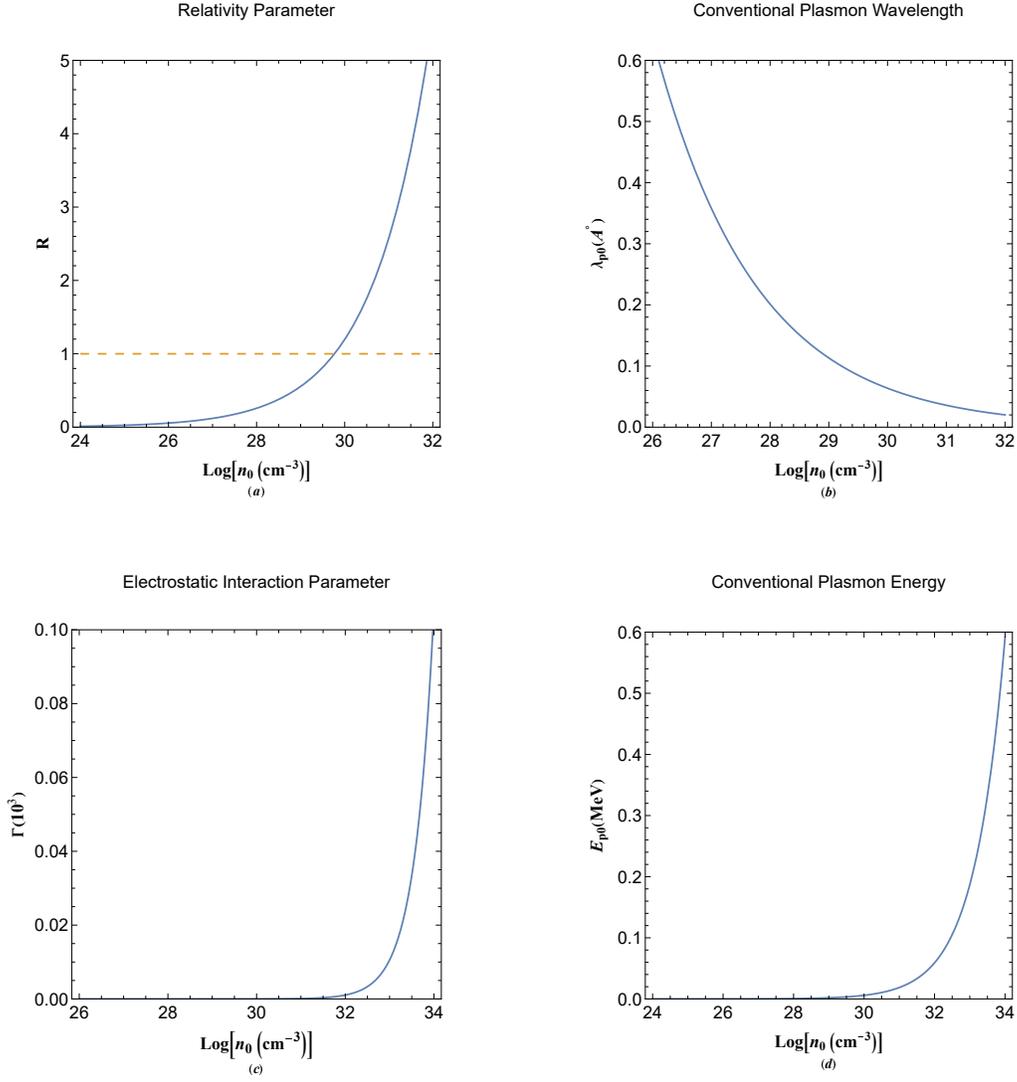}\caption{(a) variation of the relativity parameter with electron density. The dashed line indicates the value at Compton electron density. (b) Variatio of the conventional plasmon wavelength in terms of electron number density. (c) Variation of collective interaction strength parameter with electron number density. (d) Variation of conventional plasmon energy versus the electron number density.}
\end{figure}

Figure 1 shows the characteristic scale units and parameters for our normalization. The relativity parameter variation with the electron density is shown in Fig. 1(a). This parameter sharply increase with increase in the electron number density and reaches unity when the electron density coincides with the critical Compton number density $n_c\simeq 5.86\times 10^{29}$cm$^{-3}$ which is typical of white dwarfs and leads to the gravitational stellar collapse in the Chandrasekhar limit \cite{chandra}. It is well known that the electron gas pressure changes a polytropic dependence beyond this critical relativity parameter value \cite{akbcom}. Figure 1(b) depicts variation of the plasmon wavelength $2\pi/k_p$ with $k_p=\sqrt{2m_0 E_p}/\hbar$ being the plasmon wavenumber and $E_p=\hbar\omega_p$ being the plasmon energy with $\omega=\sqrt{4\pi e^2 n_0/m_0}$ being the plasmon frequency. The index zero in $\lambda_{p0}$ indicates that this parameter depends on the total equilibrium electron number density $n_0$ of the gas which is used to define conventional plasmon parameters. It will be shown that these definitions only apply in the density-temperature regimes where all electrons are excited to the so-called plasmon energy band. The conventional plasmon wavelength is seen to decrease with increase in the electron number density. The plasmon wavelength at the critical Compton density can be as low as, $\lambda_{p0}\simeq 0.0727${\AA}, which is much lower than the value of the Bohr radius, $r_B=\hbar^2/m_0 e^2\simeq 0.53${\AA}. The variation of the electrostatic interaction strength parameter $\Gamma$ is shown in Fig. 1(c). It increases sharply with electron concentration and vanishes in the single electron limit leading the system (\ref{pf}) to the original Schr\"{o}dinger equation. The variation of conventional plasmon energy with electron number density is shown in Fig. 1(d). While for metallic elements this energy varies in a few electron volts range, it increases to MeV range for fully relativistic quantum electron concentrations.

The Fourier analysis of the normalized system leads to the generalized energy dispersion $E=k^2/2+\Gamma/k^2$ in which the energy and wavenumber are normalized to the electron rest energy and Compton wavenumber, respectively. Note that in the zero electron density limit $\Gamma\to 0$ and $\mu_0\to 0$, one obtains the non-relativistic free electron dispersion $E=\hbar^2k^2/2m_0$ in dimensional units. Note that the dispersion relation is composed of two distinct branches due to both particle-like and wave-like behavior of the electron gas.

\begin{figure}[ptb]\label{Figure2}
\includegraphics[scale=0.57]{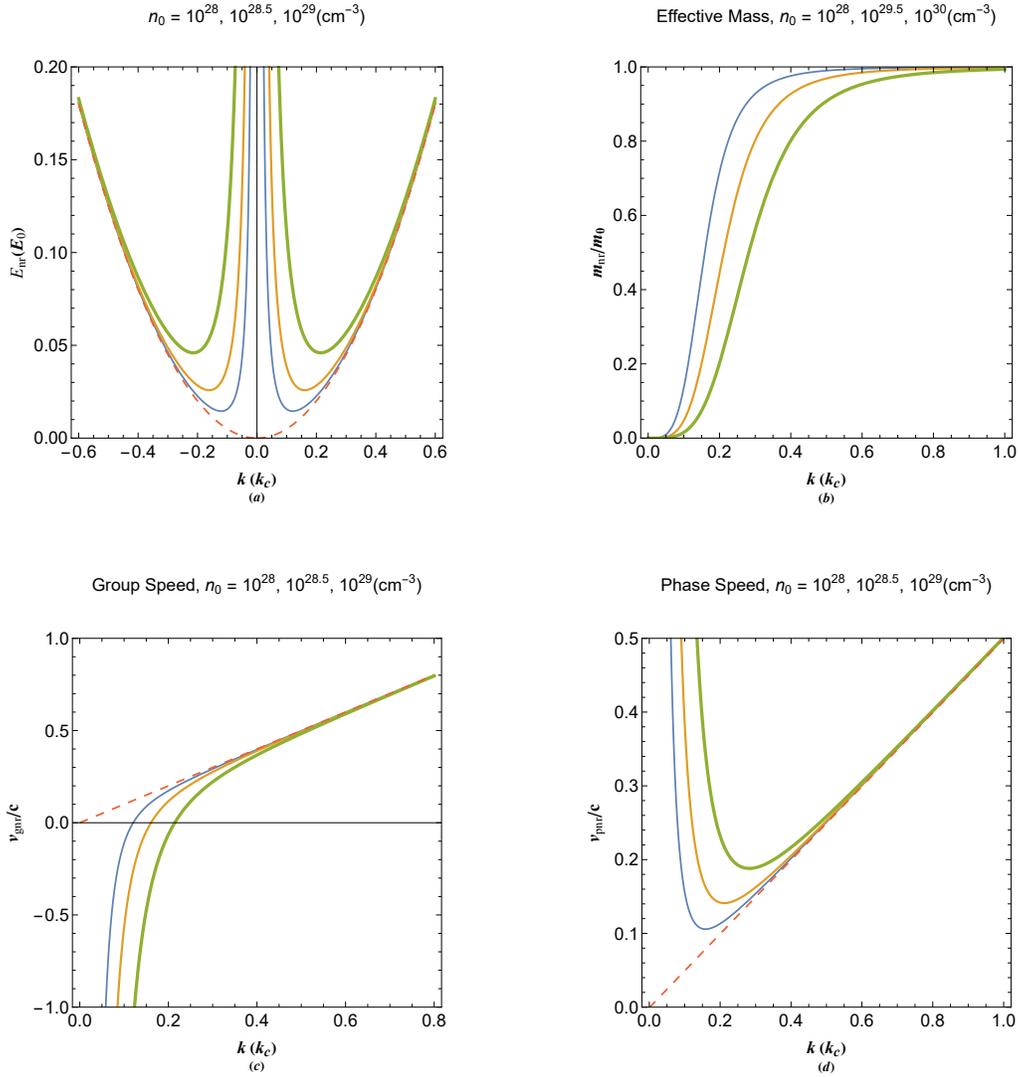}\caption{The non-relativistic plasmon energy band structure for different electron number densities. (b) The non-relativistic effective electron mass-ratio for different electron number densities. (c) The non-relativistic plasmon group speed for different electron number densities. (d) The non-relativistic phese speed of plasmon excitations for different electron number densities. The dashed curves indicates the free electron value. The increase in the thickness of curves indicate increase in the value of varied parameter above each panel.}
\end{figure}

Figure 2(a) depicts the energy band structure of non-relativistic electron gas for different electron number density in relativistic quantum regime. The dashed curve shows the non-relativistic free electron dispersion curve. The dispersion curve for a given electron density is composed of two branches connecting at the minimum plasmon conduction energy. Note that below this energy, collective quantum excitations become unstable. This feature is analogous to the features in energy band structure of crystalline solids in which electrons with energies below the conduction level do not contribute to the single electron-hole excitation phenomena, such as the electronic transport. In our case, however, electrons at the Fermi level need enough energies to excite to the plasmon conduction band in order to take part in various collective phenomena. Note that occurrence of energy band gap is due to the electrostatic interaction and vanished in the single electron limit. It is revealed that the plasmon band gap increases with in crease in the electron number density and the plasmon conduction wavenumber moves to higher values. Figure 2(b) shows the variation of non-relativistic effective electron mass ratio to the rest mass for different electron concentration. It is noted that the effective mass ratio is vanishingly low for long wavelength collective excitations and reaches the rest mass limit at the small wavelength limit. It is also remarked that effective mass decreases with increase in the electron number density of Fermi gas. The appropriately normalized fractional electron mass parameter is given as
\begin{equation}\label{m}
\frac{m}{{{m_0}}} = {\left( {\frac{{{d^2}E}}{{d{k^2}}}} \right)^{ - 1}} = \frac{{{k^4}}}{{{k^4} + 6\Gamma }},
\end{equation}
where, the single electron limit $\Gamma=0$ reduces to $m=m_0$. Variation of the normalized group speed (to the speed of light in vacuum) $v_g=dE/dk=k-2\Gamma/k^3$, of collective plasmon excitations for different electron density is shown in Fig. 2(c). The dashed line represents the free electron (zero density) group speed limit. It is remarked that group speed varies from negative to positive values unboundedly approaching the free electron value in small wavelength limit. It is also noted that increase in electron concentration lowers the group speed of collective electron excitations. The phase speed is shown in Fig. 2(d) with the dashed line corresponding to the single electrn limit. The phase speed has a minimum value for given electron number density and approached the dashed line in large wavenumber excitation limit. It is revealed that the phase speed increases by increase in electron density with its minimum value moved to lower wavelengths. The unboundedness of the collective wave speed in by no means violation of the special relativity which puts a limit on the single electron speed even in current non-relativistic model.

\section{Relativistic Quantum Energy Dispersion}

In order to study collective quantum electron excitations in the relativistic gas we use the relativistic energy dispersion $\varepsilon=\sqrt{E_0^2+p^2c^2}$ in which $p$ is the relativistic electron momentum. Consider the Hamiltonian ${\cal H}=K+E_0-e\phi({\bf r})+\mu$ in which $K$ denotes the relativistic kinetic energy. Using the identity ${\cal H}{\cal N}({\bf{r}},t)=\varepsilon{\cal N}({\bf{r}},t)$ and applying the quantum operators, $E\to i\hbar\partial/\partial t$ and $p\to -i\hbar \nabla$, we arrive at the following square-root Klein-Gordon system \cite{fhaas2}
\begin{subequations}\label{spr}
\begin{align}
&i\hbar \frac{{\partial {\cal N}({\bf{r}},t)}}{{\partial t}} = \left( {\sqrt {E_0^2 - {\hbar ^2}\Delta } } \right){\cal N}({\bf{r}},t) + e\phi ({\bf{r}}){\cal N}({\bf{r}},t) - \mu {\cal N}({\bf{r}},t),\\
&\Delta \phi ({\bf{r}}) = 4\pi e\left[ {|{\cal N}({\bf{r}}){|^2} - {n_0}} \right].
\end{align}
\end{subequations}
The isothermal EoS of a relativistic quantum electron gas can be expressed as
\begin{subequations}\label{npr}
\begin{align}
&n_e(\eta ,\zeta) = 8\pi \sqrt 2 \frac{{m_0^3{c^3}}}{{{h^3}}}{\zeta ^{3/2}}\left[ {{F_{1/2}}\left( {\eta ,\zeta} \right) + \left( {\zeta/2} \right){F_{3/2}}\left( {\eta ,\zeta } \right)} \right],\\
&P_e(\eta,\zeta) = \frac{{16\pi \sqrt 2 }}{3}\frac{{m_0^4{c^5}}}{{{h^3}}}{\zeta ^{5/2}}\left[ {{F_{3/2}}\left( {\eta ,\zeta} \right) + \left( {\zeta/2} \right){F_{5/2}}\left( {\eta ,\zeta} \right)} \right],\\
&{U_e}(\eta ,\zeta ) = 8\pi \sqrt 2 \frac{{m_0^4{c^5}}}{{{h^3}}}{\zeta ^{5/2}}\left[ {{F_{3/2}}\left( {\eta ,\zeta } \right) + \zeta {F_{5/2}}\left( {\eta ,\zeta } \right)} \right],
\end{align}
\end{subequations}
where $\eta=\mu/E_0$, $\zeta=k_B T/E_0$ and $F_k$ is the Fermi-Dirac integral defined as
\begin{equation}\label{fdi}
{F_k}\left( {\eta ,\zeta} \right) = \int\limits_0^\infty  {\frac{{{x^k}\sqrt {1 + \zeta x/2} }}{{\exp \left( {\eta + \zeta} \right) + 1}}} dx.
\end{equation}
Note also that the thermodynamic identity $n_e\nabla\mu=\nabla P_e(n_e)$ is also satisfied in this case. Using the similar separation of variables technique as in the non-relativistic case, and normalizing the linear system, one arrives at the relativistic pseudoforce system
\begin{subequations}\label{pfr}
\begin{align}
&i\hbar \frac{{d\psi(t)}}{{dt}} = \varepsilon\psi(t),\\
&\left( {\sqrt {1 - \Delta } } \right)\Psi ({\bf{r}}) + \Phi ({\bf{r}}) + E\Psi ({\bf{r}}) = 0,\\
&\Delta \Phi ({\bf{r}}) - \Gamma \Psi ({\bf{r}}) = 0.
\end{align}
\end{subequations}
Due to the asymmetric operation on space and time, the square-root Klein-Gordon system can not be solved analytically. However, the collective electrostatic wave dispersion in Wigner-Poisson system has been recently studied using this system \cite{fhaas}. The energy dispersion of the coupled system (\ref{pfr}) can be readily obtained by the series expansion of the momentum functional $f({\bf p})=\sqrt{1+p^2}$ prior to the replacement $p\to -i\hbar \nabla$ and the Fourier analysis and recollection of terms. We then obtain the eigenvalue system
\begin{equation}\label{ev}
\left( {\begin{array}{*{20}{c}}
{\sqrt {1 + {k^2}}  - E}&{ - 1}\\
\Gamma &{{k^2}}
\end{array}} \right)\left( {\begin{array}{*{20}{c}}
\Psi \\
\Phi
\end{array}} \right) = \left( {\begin{array}{*{20}{c}}
0\\
0
\end{array}} \right),
\end{equation}
which easily leads to the generalized dispersion $E=\sqrt{1+k^2}+\Gamma/k^2$ with the energy and wavenumber normalized to the rest electron energy and the Compton wavenumber, respectively. In the relativistic single electron limit the dispersion reduces to $E=\sqrt{1+k^2}$. The small wavenumber expansion of the normalized relativistic kinetic energy dispersion, i.e., $K=E-1$, leads to
\begin{equation}\label{rd}
K = \frac{{{k^2}}}{2} + \frac{\Gamma }{{{k^2}}} - \frac{{{k^4}}}{8} + O{\left( k \right)^5},
\end{equation}
which clearly coincides with the non-relativistic energy dispersion in lowest orders.

\begin{figure}[ptb]\label{Figure3}
\includegraphics[scale=0.57]{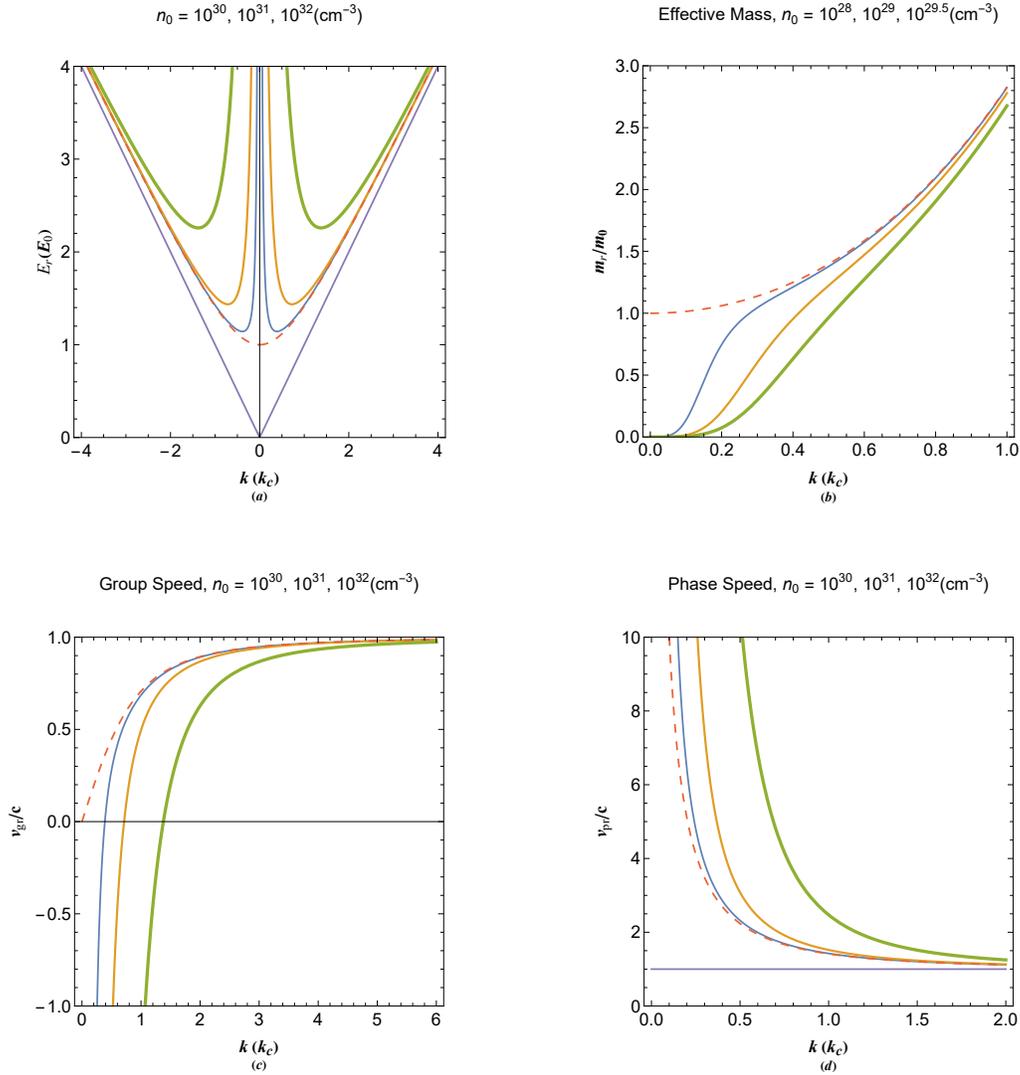}\caption{The relativistic plasmon energy band structure for different electron number densities. (b) The relativistic effective electron mass-ratio for different electron number densities. (c) The relativistic plasmon group speed for different electron number densities. (d) The relativistic phese speed of plasmon excitations for different electron number densities. The dashed curves indicates the free electron value. The increase in the thickness of curves indicate increase in the value of varied parameter above each panel.}
\end{figure}

Figure 3(a) shows the relativistic plasmon energy band structure with the dashed curve corresponding to the relativistic single electron energy dispersion. The photon lines, $E=pc$, are asymptotes to the dispersion curves of particle-like branches. It is remarked that there is an energy band gap below the plasmon conduction energy which increases with increase in electron concentration. Comparing this plot with Fig. 2(a) reveals that the increase of energy gap with electron density is more significant in the case of the relativistic quantum excitations. The variation of the relativistic mass ratio is depicted in Fig. 3(b) for different electron concentrations. The fractional mass is given as
\begin{equation}\label{mr}
\frac{m}{{{m_0}}} = {\left( {\frac{{{d^2}E}}{{d{k^2}}}} \right)^{ - 1}} = \frac{{{k^4}{{\left( {1 + {k^2}} \right)}^{3/2}}}}{{{k^4} + 6\Gamma {{\left( {1 + {k^2}} \right)}^{3/2}}}},
\end{equation}
The dashed curve corresponds to the relativistic mas of a single electron. It is seen that the fractional mass starts from zero at the long wavelength collective excitations to the infinity approaching the single electron limit at large wavenumber regime. It is also remarked that the fractional relativistic mass decreases with increase of electron density in the gas. The relativistic group speed is shown in Fig. 3(c) with the dashed curve indicating the relativistic single electron case.
\begin{equation}\label{gsr}
{v_{gr}} = \frac{{dE}}{{dk}} = \frac{k}{{\sqrt {1 + {k^2}} }} - \frac{{2\Gamma }}{{{k^3}}},
\end{equation}
The group speed increases with increase of the excitation wavenumber and reaches the speed of light in vacuum at small wavenumber limit. Finally, the phase speed decreases from infinity to the ultra-relativistic limit at small wavelength limit. The phase speed is seen to have larger values at large electron concentrations.

The collective quantum statistical behavior of electron gas may be studied using the similar procedure as developed for electron excitations. The normalized thermodynamic quntities follow
\begin{subequations}\label{ther}
\begin{align}
&{n_p}(\zeta ,\Gamma ) = \int\limits_0^\infty  {\frac{{g(k,\Gamma){d_k}E(k,\Gamma)dk}}{{1 + \exp [E(k,\Gamma)/\zeta ]}}},\\
&U_p(\zeta,\Gamma) = \int\limits_0^\infty  {\frac{{{g(k,\Gamma)}{E(k,\Gamma)}{d_k}{E(k,\Gamma)}dk}}{{1 + \exp [{E(k,\Gamma)}/\zeta]}}},\\
&C_p(\zeta ,\Gamma) = \int\limits_0^\infty  {g(k,\Gamma)E(k,\Gamma){d_\zeta }F(k,\Gamma,\zeta){d_k}E(k,\Gamma)dk},\\
&P_p(\zeta ,\Gamma ) = \int\limits_0^\infty  {\frac{{g(k,\Gamma ){d_k}E(k,\Gamma )\sqrt {{E^2}(k,\Gamma ) - 1} }}{{1 + \exp [E(k,\Gamma )/\zeta ]}}} dk,
\end{align}
\end{subequations}
where $v_g$ is the group speed of collective excitations, $n_p$, $U_p$, $C_p$ and $P_p$, respectively, denote the effective plasmon electron number density, internal plasmon energy, heat capacity of collective excitations and plasmon quantum pressure. The parameter $\zeta=T/T_0$ denotes the normalized temperature with $T_0=E_0/k_B$ is the electron rest temperature. The plasmon band density of states (DoS) is $g=(dN/dk)/|dE/dk|$ with $N=4\pi k^3/3$ being the number of plasmon modes within the spherical wavenumber volume. The DoS of non-relativistic excitations follow
\begin{equation}\label{gnr}
{g_{nr}}(k,\Gamma ) = \frac{{4\pi {k^5}}}{{\left| {{k^4} - 2\Gamma } \right|}},
\end{equation}
and for the relativistic excitations we have
\begin{equation}\label{gr}
{g_r}(k,\Gamma ) = \frac{{4\pi {k^5}\sqrt {1 + {k^2}} }}{{\left| {{k^4} - 2\Gamma \sqrt {1 + {k^2}} } \right|}},
\end{equation}

\begin{figure}[ptb]\label{Figure4}
\includegraphics[scale=0.57]{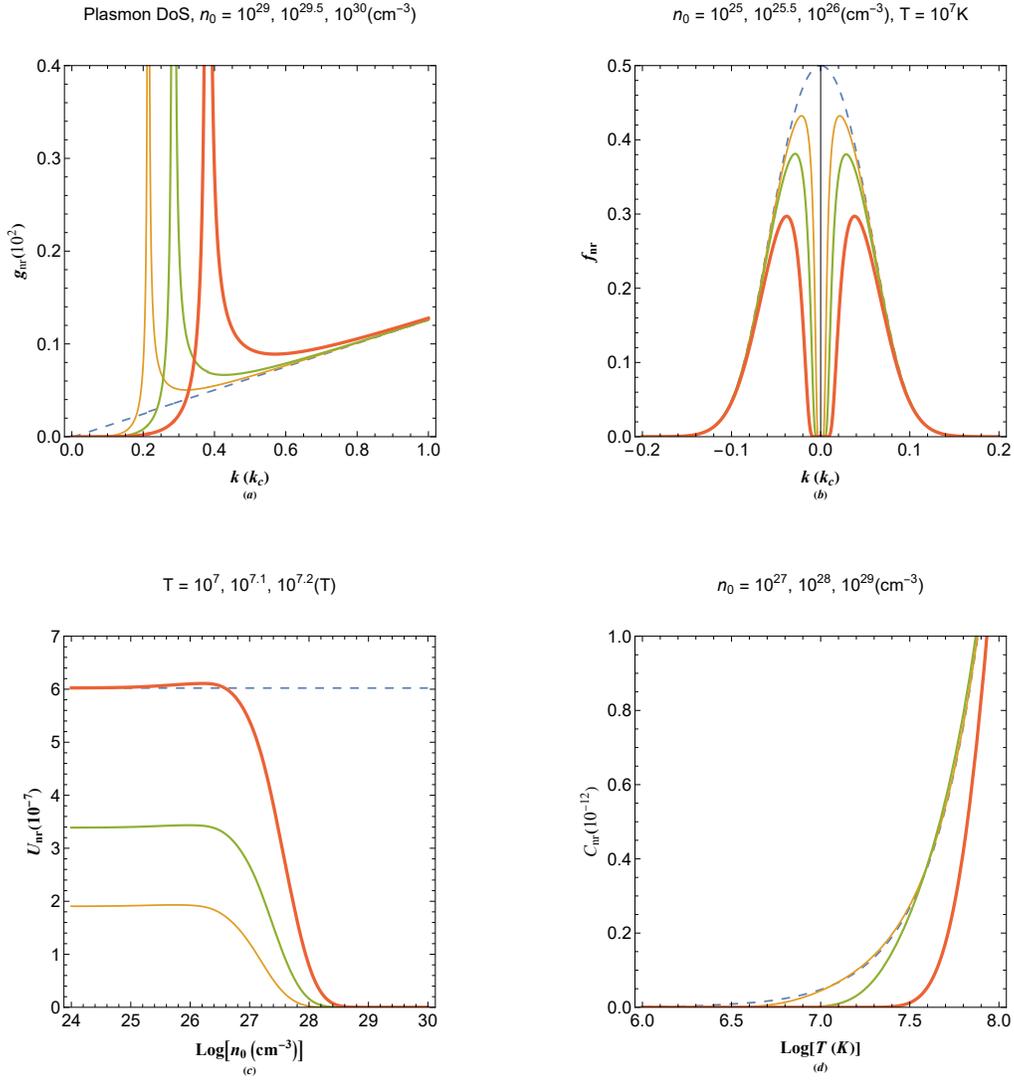}\caption{(a) The non-relativistic plasmon density of states (DoS) for various electron concentrations. (b) The non-relativistic plasmon band occupation function for various electron concentrations. (c) The non-relativistic plasmon gas internal energy variations with electron number density for different electron gas temperatures. (d) The non-relativistic plasmon gas thermal capacity variations with electron gas temperature for different electron number density. The dashed curves indicates the free electron value. The increase in the thickness of curves indicate increase in the value of varied parameter above each panel.}
\end{figure}

Figure 4(a) shows the DoS for non-relativistic plasmon excitations. The Van-Hove singularity is present due to coupling between the single-electron and collective electrostatic interactions. The DoS vanishes at the long wavelength limit and approaches the single-electron limit (dashed line) for large wavenumbers. The singularity moves to large wavenumbers with increase in the electron density. Also the density of states increase with increase of electron concentration in the gas. The non-relativistic plasmon occupation function is depicted in Fig. 4(b) for different electron number density. The dashed curve shows the occupation for single electron states. It is seen that the occupation of states with long wavelength collective excitations is strongly limited in the electron gas and the occupation probability in the non-relativistic gas decreases with increase in electron density. The internal energy of collective non-relativistic quasiparticle excitations is depicted in Fig. 4(c). The dashed line denotes the energy of free electron gas which is independent of the electron concentration. It is indeed shown that the internal energy of plasmon excitations depends on the electron concentration and vanishes at some critical electron density. This quantity increases with increase of electron gas temperature. The later feature is a unique behavior of collective expiations in the electron gas and leads to the plasmon black-out effect at high electron densities. The thermal capacity of collective excitations is shown in Fig. 4(d) indicating an increase in this quantity with increase of electron concentration. The free electron gas heat capacity is shown as dashed curve. At very high electron density the thermal capacity of a non relativistic plasmon approaches that of the free electron gas.

\begin{figure}[ptb]\label{Figure5}
\includegraphics[scale=0.57]{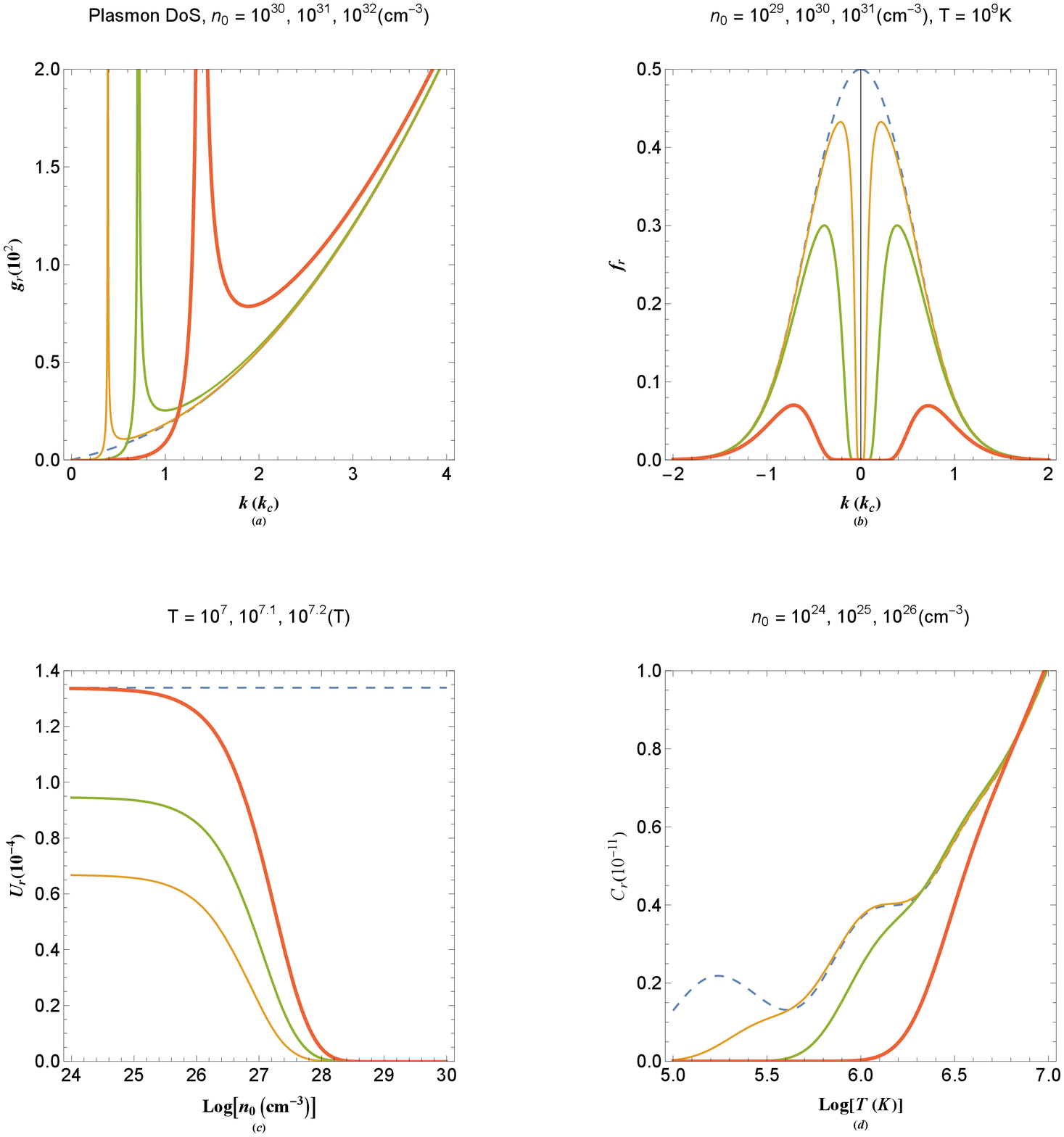}\caption{(a) The relativistic plasmon density of states (DoS) for various electron concentrations. (b) The relativistic plasmon band occupation function for various electron concentrations. (c) The relativistic plasmon gas internal energy variations with electron number density for different electron gas temperatures. (d) The relativistic plasmon gas thermal capacity variations with electron gas temperature for different electron number density. The dashed curves indicates the free electron value. The increase in the thickness of curves indicate increase in the value of varied parameter above each panel.}
\end{figure}

Figure 5(a) shows the DoS of relativistic electron gas with the dashed curve being the relativistic free electron gas DoS. It is seen that the Van-Hove-like singularity is present in the relativistic case, as well. The increase in DoS with increase of electron density more profound inthis case as compared to the non-relativistic DoS in Fig. 4(a). Figure 5(b) shows that the occupation of relativistic gas also drops at long wavelength excitations but approached the relativistic single electron curve (dashed curve) at short wavelength limit. The increase of the electron density leads relatively sharp drop of collective quasiparticle occupation probability. Figure 5(c) shows the internal plasmon energy of the relativistic electron gas with the dashed line denoting the value for relativistic free electron gas. It is seen that the internal energy in this case also drops to zero at a critical electron density. The internal energy increases with increase of electron gas temperature. The thermal capacity of relativistic interacting electron gas is shown in Fig. 5(d) with the dashed curve denoting the heat capacity of relativistic non-interacting electron gas. The heat capacity increases sharply with increase of temperature and approaches the free electron curve which shows an oscillatory behavior. Note that for given electron gas temperature the plasmon heat capacity is lower for higher electron number density.

The effective plasmon electron number density is given as
\begin{equation}\label{np}
{n_p}\left( {T,n} \right) = \int\limits_0^\infty  {\frac{{4\pi {k^2}dk}}{{1 + \exp [E(k,n)/\zeta \left( T \right)]}}} /\int\limits_0^\infty  {\frac{{4\pi {k^2}dk}}{{1 + \exp [{E_e}(k)/\zeta \left( T \right)]}}},
\end{equation}
where $E$ is the plasmon energy dispersion relation and $E_e=E(\Gamma=0)$ is the corresponding free electron dispersion. Note that the maximum value of $n_p=n_0$ corresponds to the case where all Fermi electrons are excited to the plasmon band. Then all the plasmon parameters may be defined in terms of this effective plasma electron number density. For instance the plasmon energy is given as $E_p(T,n_0)=\hbar\sqrt{4\pi e^2 n_p/m_0}$ and the plasmon wavelength is redefined as $\lambda_p(T,n_0)=2\pi\hbar/\sqrt{2m_0 E_p(T,n_0)}$. Note that in current definitions all plasma parameters depend on total equilibrium number density as well as the electron gas temperature.

\begin{figure}[ptb]\label{Figure6}
\includegraphics[scale=0.57]{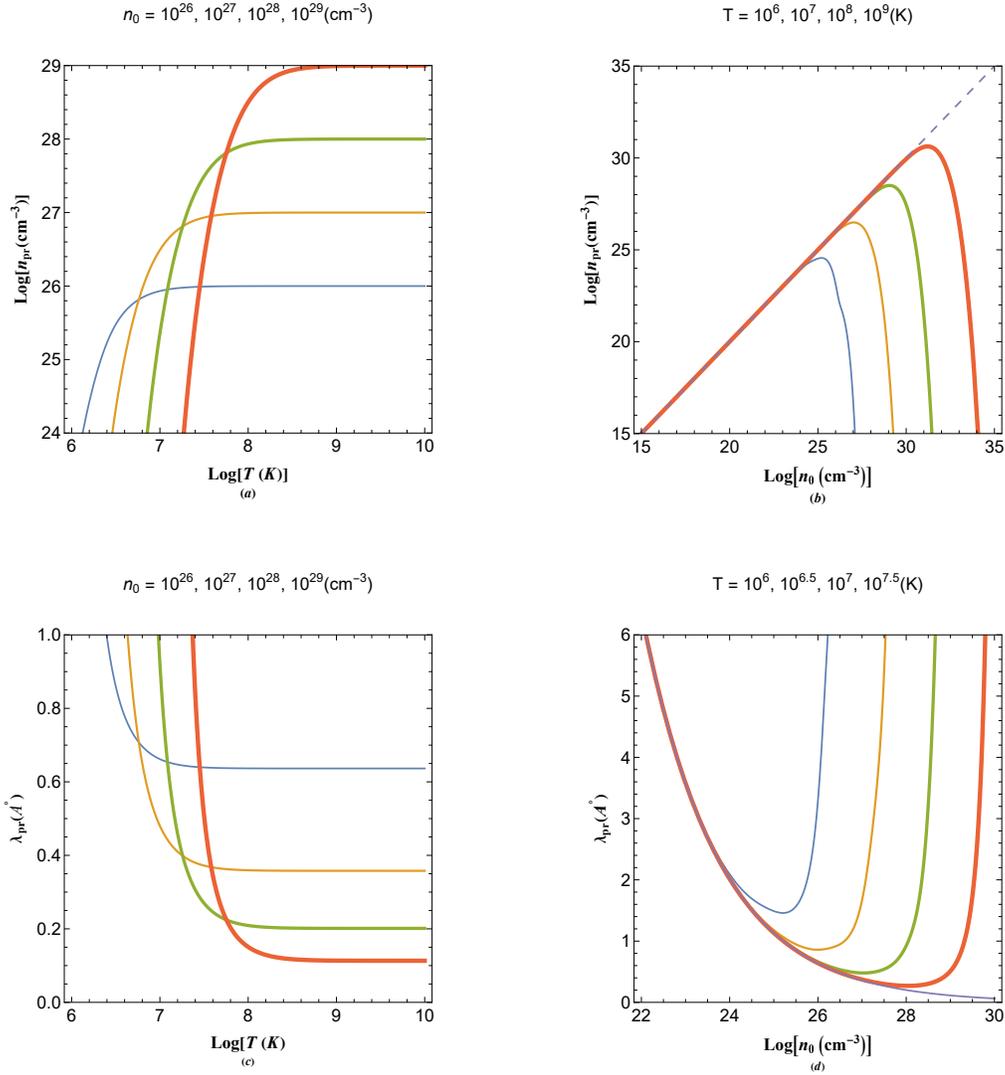}\caption{(a) Variation of plasmon number density with electron temperature for different electron number density. (b) Variation of plasmon number density with electron number density for various electron temperature. The dashed line shows the complete plasmonization limit. (c) Variation of the plasmon wavelength with electron temperature for different electron concentrations. (d) Variations of plasmon wavelength with electron number density for different electron gas temperatures. The increase in the thickness of curves indicate increase in the value of varied parameter above each panel.}
\end{figure}

Figure 6(a) show the variation of relativistic plasmon number density in terms of the electron temperature for different total electron number density. It is remarked that with increase of the temperature the plasmon density increases and reaches a saturated value which is , let say, the complete plasmonization state. It is also remarked that the electron gas with lower electron density reaches this state faster. We have shown the plasmon electron EoS in Fig. 6(b). The dashed line denotes the complete plasmonization state. It is remarked that for a given temperature up to a critical electron density the electron gas is in the sate of complete plasmonaization state and with further increase on the electron density the plasmon density sharply drops to zero value. We refer to the state of zero plasmon electron density the plasmon black-out state and takes place sooner at lower electron gas temperatures. Figure 6(c) depicts the variation of plasmon wavelength which is a characteristic length to many plasmonic and photo-plasmonic phenomena.  It is seen that the plasmon wavelength decreases with increase of electron temperature and becomes constant after the critical temperature for a given electron number density is reached. Note that the higher the electron density the lower saturated plasmon wavelength is. Figure 6(d) reveals an important feature of the electron gas. It is remarked that the plasmon wavelength has minimum value corresponding to electron number density for given electron gas temperature. This minimum plasmon wavelength shifts to larger electron number density with increase of the temperature and corresponding wavelength becomes lower.

\begin{figure}[ptb]\label{Figure7}
\includegraphics[scale=0.57]{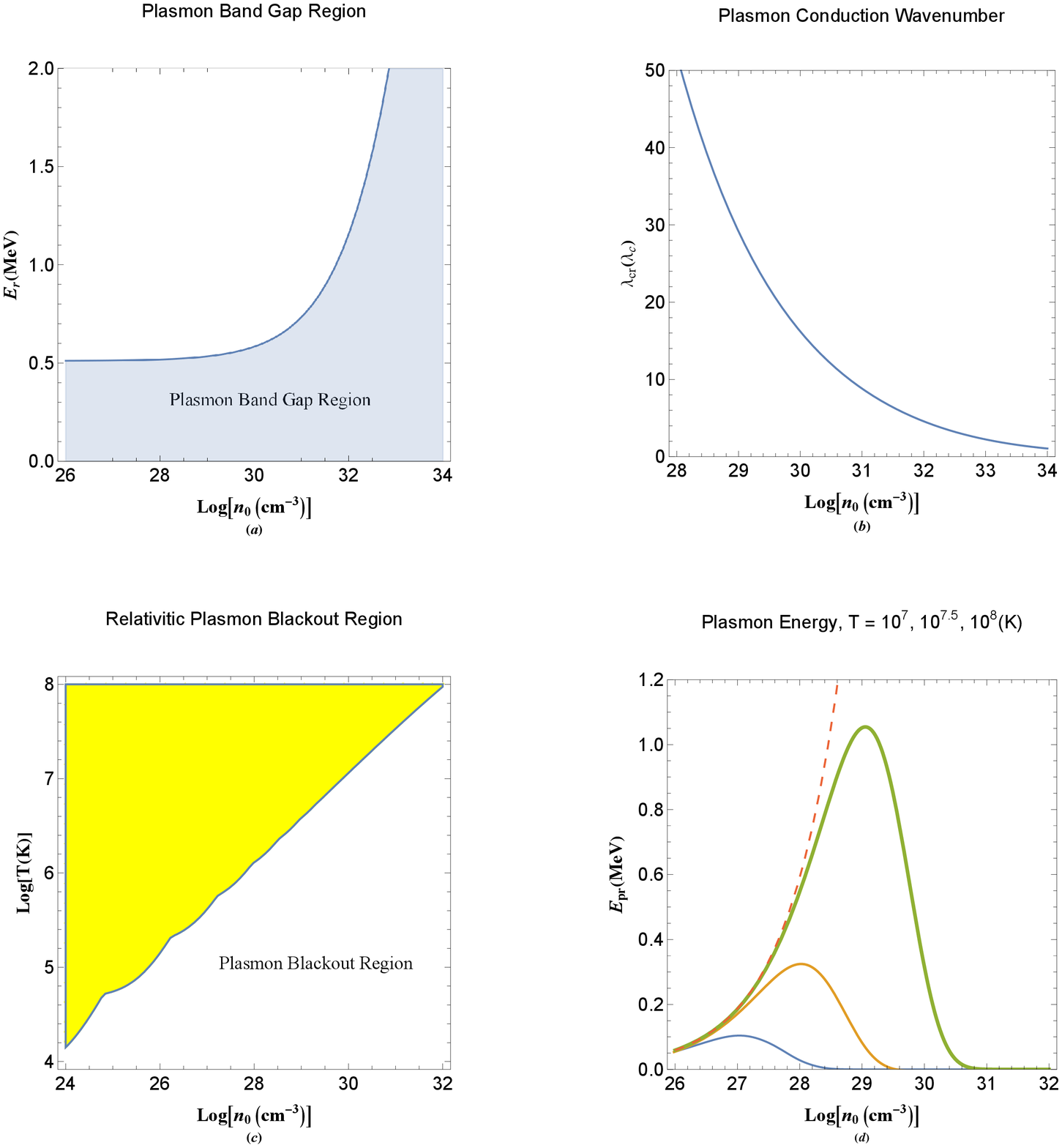}\caption{(a) Variation of relativistic plasmon energy band gap region in terms of electron number density. (b) Variation of the relativistic plasmon conduction valley wavenumber with electron concentration. (c) The relativistic plasmon black-out region in the electron temperature-density regime. (d) Variation of relativistic plasmon energy with electron number density for different values of electron temperatures. The increase in the thickness of curves indicate increase in the value of varied parameter above each panel.}
\end{figure}

Figure 7(a) shows the unstable collective excitation region due to plasmon energy band gap for a relativistic electron gas. It is noted that up to the critical density of a white dwarf star the band gap remains constant and equal to the rest energy of an electron but increases sharply for higher electron number densities. The plasmon conduction valley wavelength (normalized to the Compton wavelength) is plotted in terms of electron number density. This wavenumber decreases monotonically as the electron density increases. The plasmon black-out region in temperature-density plane is depicted in Fig. 7(c). This is the region where collective excitations become ineffective and electrons fall below the Fermi energy level (Fermi sea). The border indicates a quantum jumping feature which should be further studied and is out of the scope of current investigation. For the critical Compton electron density $n_c$ the critical temperature below which the plasmon black-out sets in is approximately, $T_c\simeq 8.8938\times10^6$K. Figure 7(d) shows the variation of plasmon energy with the electron number density for different electron gas temperatures. It is remarked that for any given temperature the plasmon energy maximizes at some electron number density which sharply shifts to higher density values and its peak value increases significantly with increase of the electron gas temperature.

\begin{figure}[ptb]\label{Figure8}
\includegraphics[scale=0.57]{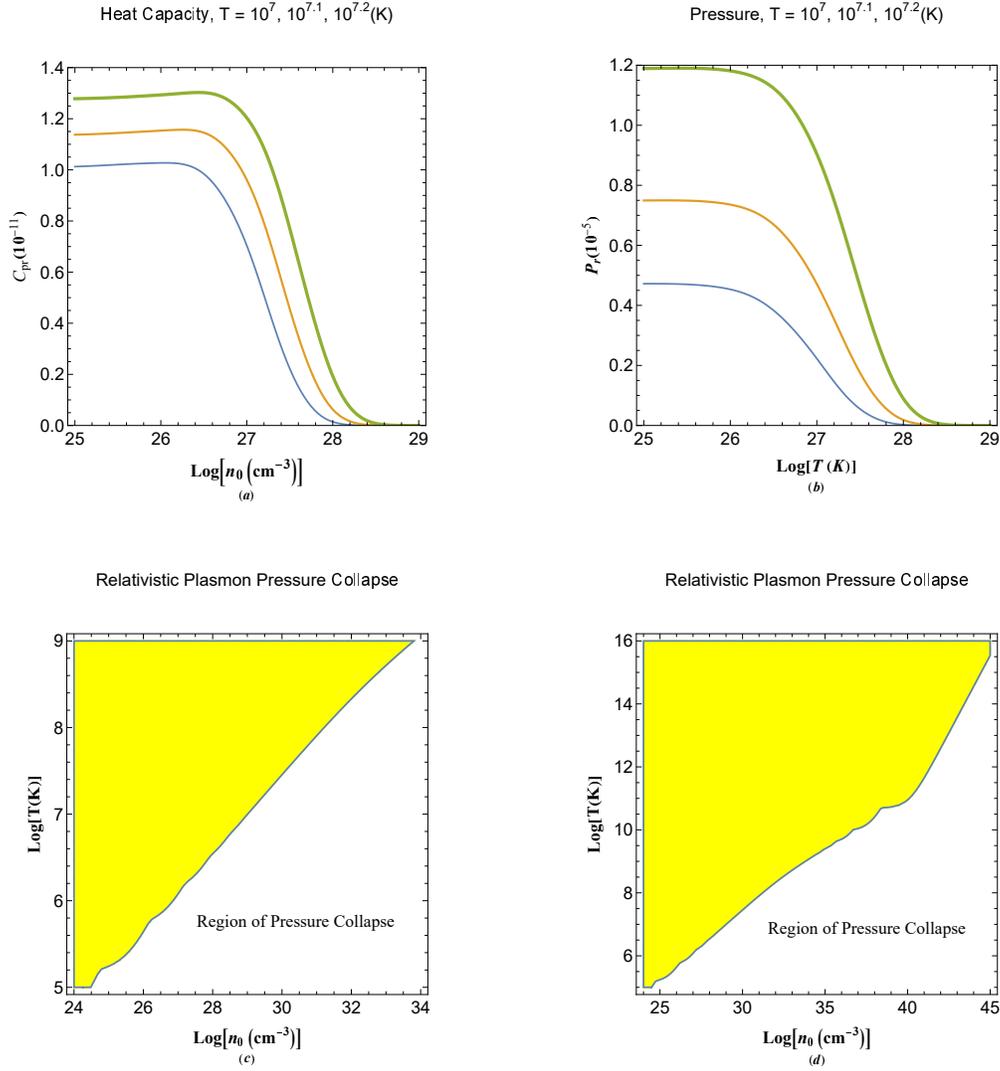}\caption{(a) The relativistic heat capacity of collective excitations in terms of electron density for different temperature values. (b) The relativistic quantum pressure of collective excitations in terms of electron temperature for different electron number density values. (c) The relativistic quantum pressure collapse region in electron density-temperature regimes. (d) The relativistic quantum pressure collapse region in electron density-temperature regimes in a wide angle. The increase in the thickness of curves indicate increase in the value of varied parameter above each panel. The increase in the thickness of curves indicate increase in the value of varied parameter above each panel.}
\end{figure}

Figure 8(a) shows the variation of the plasmon heat capacity variation with electron density for various electron temperature values using the relativistic quantum electron gas model. It is seen that the plasmon heat capacity drops to zero at a given electron temperature. The later feature is due to plasmon black-out effect, referred to earlier. However, the heat capacity cut-off takes place at relatively higher electron densities for higher electron temperatures. The plasmon pressure of relativistic electron gas is shown in Fig. 8(b). The interesting feature in this plot is the plasmon pressure collapse at a given electron density for given temperature. This is analogous to the free electron gas relativistic degeneracy pressure collapse at Chandrasekhar mass limit which is caused by the gravitational crunch \cite{chandra3}. The region of plasmon pressure collapse is shown in Fig. 8(c). It is remarkable that the collapse parameter in current study coincides with that of the Chandrasekhar coordinates, that is, $T=3.94323\times10^6$K at Compton electron number density $n_0=5.86478\times10^{29}$cm$^{-3}$, which is typical electron number density of white dwarf stars. Note that the quantum jumping feature is also present in this plot. The plasmon pressure phase region is shown in a larger scale in Fig. 8(d). It is revealed that the pressure collapse border undertakes a polytropic phase change at a very high critical electron number density and temperature. The nature of such critical behavior is not revealed at present work and needs further investigations.

\begin{figure}[ptb]\label{Figure9}
\includegraphics[scale=0.57]{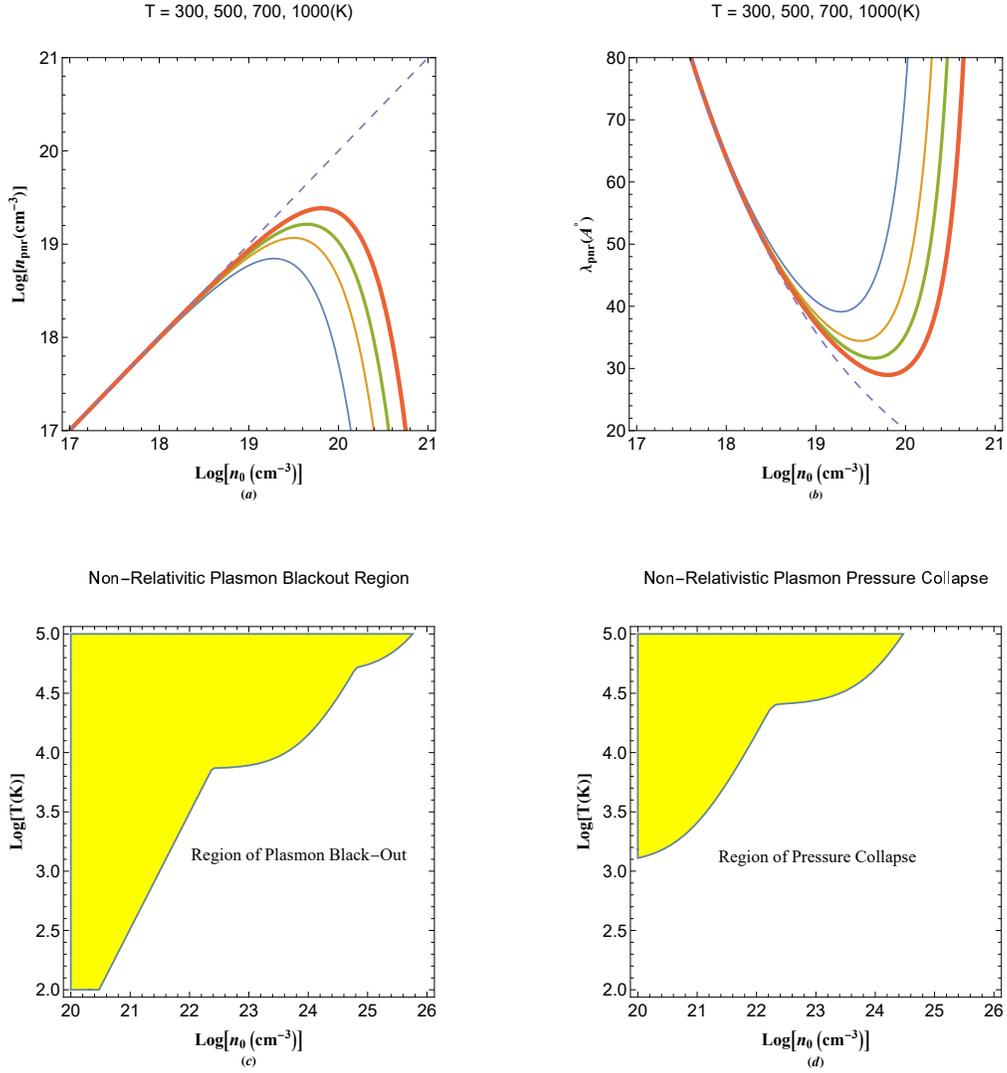}\caption{(a) The variation of non-relativistic plasmon number density as a function of total electron number density for various low temperature values. (b) The variations of non-relativistic plasmon wavelength with electron concentration for different value of electron temperatures. (c) The non-relativistic plasmon black-out region in low density-temperature regime. (d) The non-relativistic plasmon pressure collapse region in low density-temperature regime. The increase in the thickness of curves indicate increase in the value of varied parameter above each panel.}
\end{figure}

Figure 9(a) shows the plasmon number density EoS using the non-relativistic model. It is remarked that similar feature as in the relativistic electron gas is present in which the plasmon density cut-off occurs beyond a critical density for given electron temperature. For instance the plasmon cut-off electron density at room temperature is approximately $n_0=9.18765\times10^{20}$cm$^{-3}$ which is relatively lower than the typical metallic electron density. Therefore, most elemental metals reside in the plasmon black-out region. Few years ago, Glenzer et al. \cite{glenzer} have reported observations of the electron plasma oscillations in a solid density plasma with the peak electron number density around $n_0=3\times10^{23}$cm$^{-3}$ and the equilibrium electron temperatures of $12$eV ($1.4\times 10^5$K, which is relatively higher than the Fermi electron temperature for metals, by using collective X-ray scattering techniques. This shows that metals conduct plasmon at relatively higher temperatures. Our findings on plasmon black-out and plasmon pressure collapse may have a profound implications for the inertial confinement fusion (ICF), the EoS of warm dense matter (WDM) and evolution of stellar and other cosmological structures such as the mysterious dark matter. The variation of plasmon wavelength in low temperature regime is depicted in Fig. 9(b) using the non-relativistic model. The minimum plasmon wavelength at room temperature takes place at electron density of $n_0=1.9034\times10^{19}$cm$^{-3}$ corresponding to the value of $\lambda_p\simeq 3.91254$nm which resides in the ultraviolet radiation spectrum. The latter aspect of the electron gas is closely related to the surface plasmon resonance effect \cite{}. The plasmon black-out and pressure collapse regions are shown respectively in Figs. 9(c) and 9(d) in the non-relativistic quantum electron gas model. The quantum jumping feature is pronounced and doe not depend on the relativistic considerations. At room temperature the plasmon black-out takes place at critical density, $n_0=9.18765\times10^{20}$cm$^{-3}$. on the other hand, for metallic electron density $n_0\simeq 10^{22}$cm$^{-3}$ the plasmon black-out takes place below the critical temperature of $T\simeq 3110$K. The non-relativistic plasmon pressure collapse for electron number density typical of metals $n_0\simeq 10^{22}$cm$^{-3}$ occurs below temperature $T\simeq 628.251$K. Note that the degeneracy pressure is quite different from the plasmon pressure and acts effectively under very strong external forces such as the gravity in extreme environment.

\begin{figure}[ptb]\label{Figure10}
\includegraphics[scale=0.57]{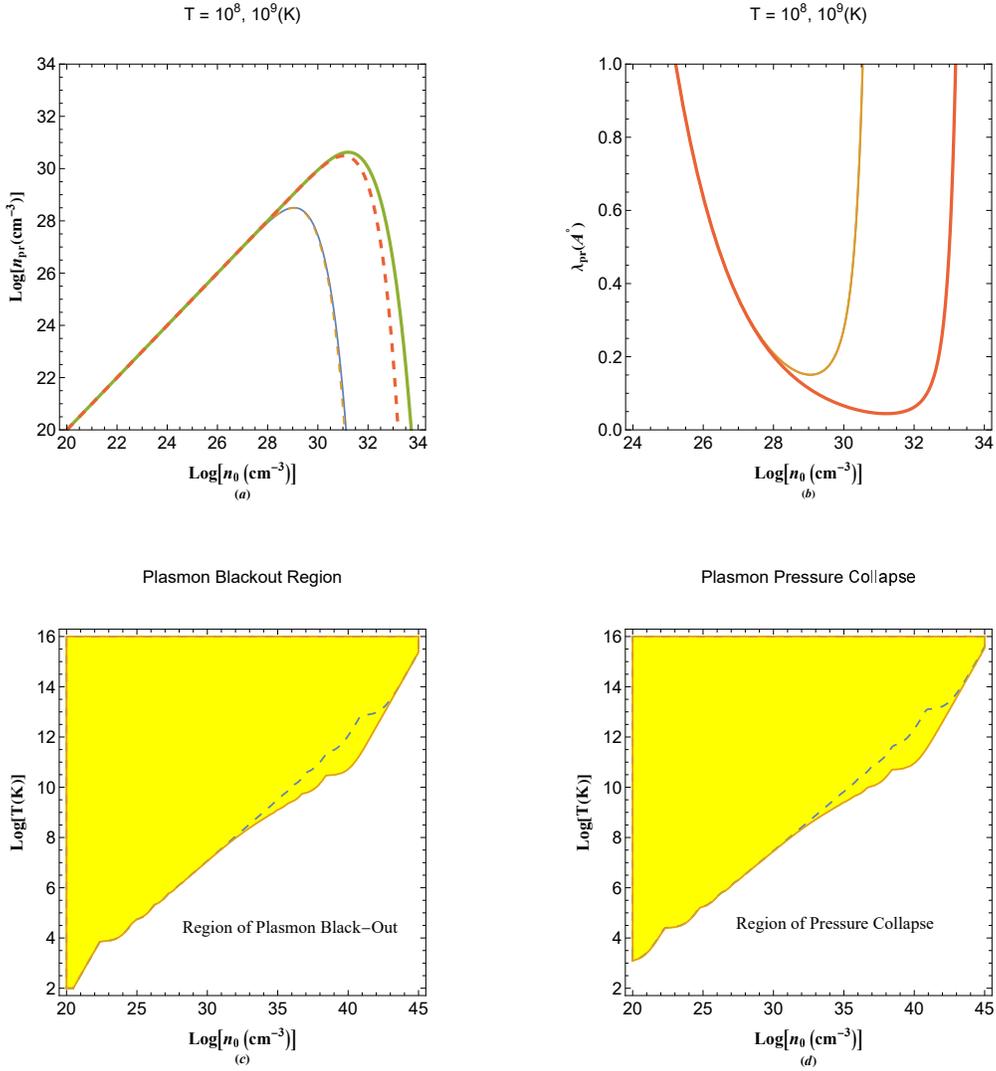}\caption{(a) Comparison of plasmon EoS curves for relativistic (solid curve) and non-relativistic (dashed curve) models for different electron temperatures. (b) Comparison of plasmon wavelengths for relativistic (solid curve) and non-relativistic (dashed curve) models for different electron temperatures. (c) Comparison of plasmon black-out regions for relativistic (solid border) and non-relativistic (dashed border) models. (d) Comparison of plasmon pressure collapse regions for relativistic (solid border) and non-relativistic (dashed border) models. The increase in the thickness of curves indicate increase in the value of varied parameter above each panel.}
\end{figure}

In Fig. 10 we compare the predictions of non-relativistic and relativistic models. The dashed/solid curves correspond to the non-relativistic/relativistic model. Figure 10(a) shows EoS for two different electron temperature. For the lower temperature $T=10^8$K the two models do not show significant difference whereas for $T=10^9$K there is slight difference. The plasmon wavelength simulation using both models predict almost the same result as shown in Fig. 10(b). It is remarked from Fig. 10(c) that deviation from relativistic effects in the non-relativistic model of plasmon black-out region only starts at density-temperature range much above that of the white dwarf stars. However for much higher electron densities the non-relativistic plasmon black-out region is larger than that for relativistic model. Figures 10(c) and 10(d) suggests that the non-relativistic model predicts relatively accurate results for the plasmon black-out and pressure collapse up to the white dwarf stars temperature-density regime.

\section{Electron-Positron Energy Band structure}

The relativistic quantum model can be extended to include the positron species. We then have the following normalized coupled linear system
\begin{subequations}\label{pfrp}
\begin{align}
&i\hbar \frac{{d\psi(t)}}{{dt}} = \varepsilon\psi(t),\\
&\left( {\sqrt {1 - \Delta } } \right){\Psi _e}({\bf{r}}) + \Phi ({\bf{r}}) + E{\Psi _e}({\bf{r}}) = 0,\\
&\left( {\sqrt {1 - \Delta } } \right){\Psi _p}({\bf{r}}) - \Phi ({\bf{r}}) + (E + 2){\Psi _p}({\bf{r}}) = 0,\\
&\Delta \Phi ({\bf{r}}) - \Gamma {\Psi _e}({\bf{r}}) + \sigma\Gamma {\Psi _e}({\bf{r}}) = 0,
\end{align}
\end{subequations}
where $\sigma=n_{p0}/n_{e0}$ is the fractional positron-to-electron density ratio and it has a dependence on the pair production rate with the ambient plasma temperature. The system (\ref{pfrp}) admits the following dispersion relation
\begin{subequations}\label{epd}
\begin{align}
&{{\rm E}_ - } = \frac{{2{k^2}\left( {\sqrt {1 + {k^2}}  - 1} \right) + \left( {1 + \sigma } \right)\Gamma  - \sqrt {4{k^4} + 4{k^2}\left( {1 - \sigma } \right)\Gamma  + {{\left( {1 + \sigma } \right)}^2}{\Gamma ^2}} }}{{2{k^2}}},\\
&{{\rm E}_ + } = \frac{{2{k^2}\left( {\sqrt {1 + {k^2}}  - 1} \right) + \left( {1 + \sigma } \right)\Gamma  + \sqrt {4{k^4} + 4{k^2}\left( {1 - \sigma } \right)\Gamma  + {{\left( {1 + \sigma } \right)}^2}{\Gamma ^2}} }}{{2{k^2}}}.
\end{align}
\end{subequations}
Unless the exact dependence of the parameter $\sigma$ is known to the temperature, current model can not predict realistic results.

\begin{figure}[ptb]\label{Figure11}
\includegraphics[scale=0.57]{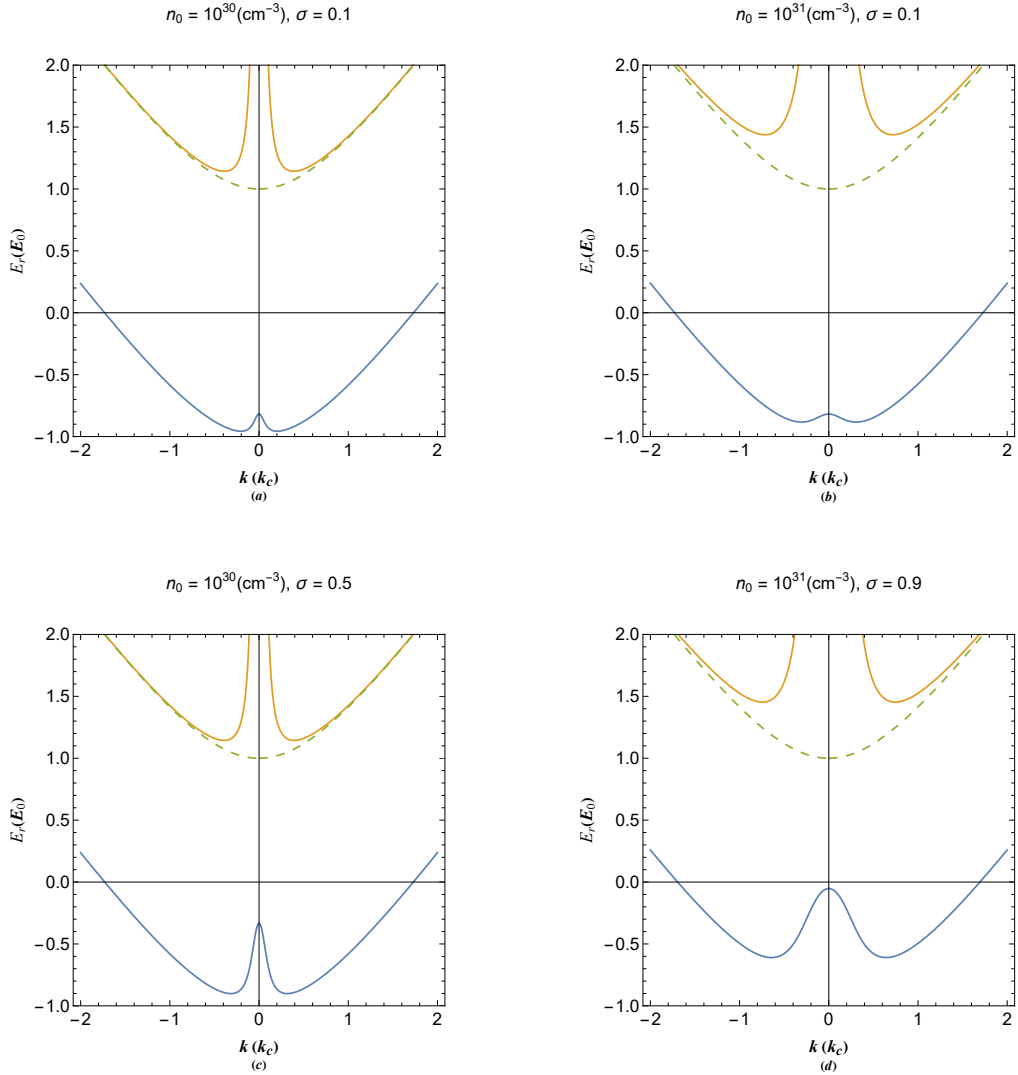}\caption{(a) The electron positron relativistic quantum plasmon excitations for $\sigma=0.1$ and $n_0=10^{30}$cm$^{-3}$. (b) The electron positron relativistic quantum plasmon excitations for $\sigma=0.1$ and $n_0=10^{31}$cm$^{-3}$. (c) The electron positron relativistic quantum plasmon excitations for $\sigma=0.5$ and $n_0=10^{30}$cm$^{-3}$. (d) The electron positron relativistic quantum plasmon excitations for $\sigma=0.9$ and $n_0=10^{31}$cm$^{-3}$. The dashed curves indicates the free electron value. }
\end{figure}

Figure 11 shows the band structure of relativistic quantum electron-positron pair plasmon excitations in terms of different parameters. The dashed curve indicates a single branch in the absence of collective interactions. It is seen in Fig. 11(a) that another dispersion branch which falls below the Fermi level appears which due to the presence of positron species. The positron band is analogous to the hole band in semiconductor band structure and the negative energies of the positron band indicates the presence of pair production much below the Fermi sea. There are stable pair plasmon excitations below the Fermi level due to the presence of positrons. However, these excitations are strongly damped at higher wavenumber values due to collisionless Landau effect. Figure 11(b) shows the bans structure profile at elevated electron density. It is remarked that the positron band departs from the bottom of Fermi sea leading to the increased positron energies. Moreover, Fig. 11(c) reveals that increase in fractional positron density strongly affects the positron energies at large wavelength limit, increasing the energy of collective positron oscillations. Finally, Fig. 11(d) shows that further increase of electron density leads to elevation of both negative and positive dispersion branch energies shifting the positron plasmon conduction valley to higher wavenumber values.

\section{Conclusion}

We studied the effect of collective energy band structure on various thermodynamic parameters in the framework of non-relativistic and relativistic quantum models. The effective Schr\"{o}dinger-Poisson and square-root Klein-Gordon-Poisson models are Fourier analyzed and the energy band structure describing the collective oscillation in an electron gas of arbitrary degeneracy is obtained. Both models predict novel features of plasmon black-out and pressure collapse due to the plasma electron density cut-off at large density and low temperatures. The later is because electron excitation probability to excite beyond the plasmon band gap reduces at high density and low temperature regimes. Using the same model we studies the influence of positron species on the energy band structure which show appearance of low lying distinct positron band below the Fermi electron sea. Current findings may direct consequences for inertial confinement scheme and equation of state (EoS) of warm dense matter (WDM).

\section{Data Availability}

The data that support the findings of this study are available from the corresponding author upon reasonable request.

\end{document}